\documentclass[aps,prl,twocolumn,superscriptaddress,showpacs]{revtex4}
\usepackage{graphicx}
\usepackage{amsfonts}
\usepackage{amssymb}
\usepackage{array}
\usepackage{amsmath}
\usepackage{color}
\usepackage{verbatim} 
\newcommand{\ket}[1]{\left\vert#1\right\rangle}

\newcommand{\pjct}[2]{\left\vert#1\right\rangle\!\left\langle#2\right\vert}
\begin{document}
\title{Production of entanglement with highly-mixed states
}

\author{Minsu Kang}
\affiliation{Center for Macroscopic Quantum Control,  Department of Physics and Astronomy,
Seoul National University, Seoul, 151-742, Korea}

\author{M. S. Kim}
\affiliation{QOLS, Blackett Laboratory, Imperial College London, London SW7 2BZ, United Kingdom}

\author{Hyunseok Jeong}
\affiliation{Center for Macroscopic Quantum Control, 
Department of Physics and Astronomy,
Seoul National University, Seoul, 151-742, Korea}
\date{\today}

\begin{abstract}
We study production of entanglement with highly-mixed states.
We find that entanglement between highly mixed states can be generated 
via a direct unitary interaction even when both states have purities arbitrarily close to zero.
This indicates that purity of a subsystem is not required for entanglement generation.
Our result is in contrast to previous studies 
where the importance of the subsystem purity was emphasized.
\end{abstract}

\maketitle

Entanglement is considered a genuine quantum correlation that cannot be described by any
classical means. 
Meanwhile,  thermal states, particularly when they are in heavy mixtures, are regarded as classical states.
 In general, generating entanglement using 
a classical state is much more difficult than when using a nonclassical state.
For example, it was shown that nonclassicality is a prerequisite of generating entanglement of light fields using a beam splitter \cite{SonPRA2002}.

On the other hand, it is possible to generate entanglement with highly mixed thermal states under certain conditions 
\cite{Filip2001,BosePRL2001,KimPRA(R)2002,Fer2006,JeongPRL2006}.
Bose \textit{et al.} showed \cite{BosePRL2001} that entanglement always arises between a two-level atom and a thermal field inside a cavity
irrespective of the temperature of the thermal state as long as the atom was
initially in a pure excited state.
In their work, the Jaynes-Cummings (JC) interaction was used to model the cavity interaction.
This indicates that the purity of the atom enforces the atom and the thermal state to become entangled
even when the thermal state is extremely mixed.
Kim {\it et al.} showed  \cite{KimPRA(R)2002}  that two atoms can become entangled through their interactions with a thermal field
when the two atoms are initially in their pure states.
Jeong and Ralph also pointed out \cite{JeongPRL2006} that
entanglement between thermal states at arbitrarily high temperatures can be generated
using a cross-Kerr nonlinear interaction if an ancillary microscopic superposition is used with a conditioning measurement.

Indeed, these are not the end of the investigations.
For example, another interesting question would be whether 
a thermal state at an arbitrarily high temperature can ever be entangled with a mixed atomic state
by a direct unitary interaction.
It also remains unanswered whether entanglement may be generated between
thermal states at arbitrarily high temperatures by a direct unitary interaction.
In this paper, we study several examples to answer these questions.
We conclude that entanglement between two highly mixed states can be generated 
solely by a unitary interaction even when 
both states have purities arbitrarily close to zero.
Our results show that subsystem purity is not a necessary condition
for generating entanglement by a direct unitary interaction.
Thus the importance of the initial purity in entanglement generation depends on the interaction model.

\begin{figure}[t]
\includegraphics[width=245px]{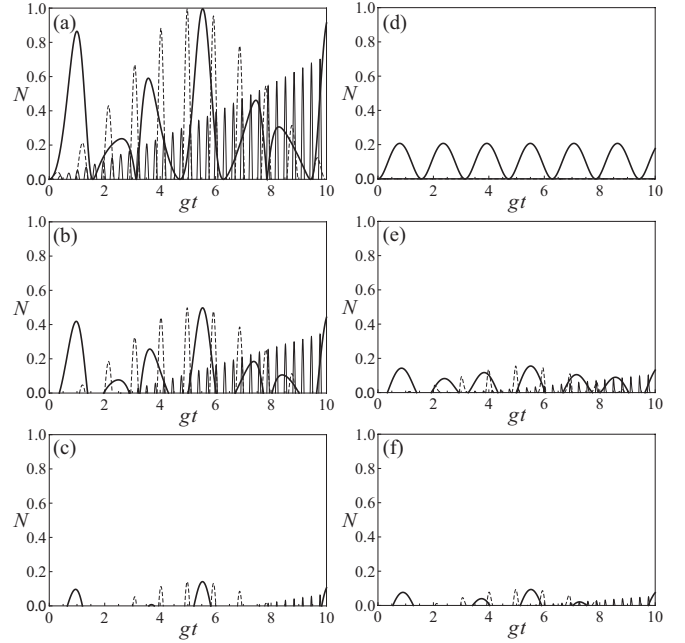}
\caption{Negativity of partial transpose (NPT) of the projected density matrix in Eq.~(\ref{eq:BK}) versus (interaction strength $g$)$\times$(interaction time $t$)
 for entanglement between an atom and a thermal state discussed in the text.
  The projection basis on the field mode is $\{\pjct{n}{n},\pjct{n+1}{n+1}\}$ with $n=0$ (thick), $n=10$ (dashed), and $n=100$ (thin). The temperature of the field is assumed to be infinite for (a-c) 
and purity $p$ of the atomic state is (a) 1, (b) 0.9, and (c) 0.8.  The atom is assumed to be maximally mixed ($p=1/2$) for (d-f) and $\lambda$ of the field is (d) 0, (e) 0.1, and (f) 0.2 in order of decreasing purity. 
}
\label{fig:BK}
\end{figure}

Reference \cite{BosePRL2001} shows that entanglement is always generated from a pure excited atom and a
thermal field through a JC interaction irrespective of the temperature of the field.
If both the atomic state and the cavity field are in thermal states with a same temperature, entanglement disappears in the infinite temperature
limit \cite{BosePRL2001}.
The reason for this is that the total state becomes proportional to the identity as the temperature goes to infinity \cite{BosePRL2001}.
However, it is possible to prepare the initial atomic state in an independent manner from
the temperature of the field.
We here examine such examples where 
both parties are in mixed states but their degrees of purities are independently varied.

We first consider an atomic state, $p \pjct{e}{e} + (1-p) \pjct{g}{g}$ 
with $0\leq p\leq1$, and a thermal-field state, $\rho^{th} = (1-\lambda)\sum_n \lambda^n 
\pjct{n}{n}$, where $\ket{g}(\ket{e})$ is the ground (excited) state of the atom and $\ket{n}$ is the photon number state of the field. We also note that
$\lambda = \textnormal{exp}[-\hbar\omega/k_B T]$, $k_B$ is the Boltzman constant, $T$  is the temperature, $\hbar$ is the Plank constant, and
$\omega$ is the frequency of the optical field.
In our analysis, the purity of state $\rho$ is quantified by the linear entropy $\textnormal{Tr}[\rho^2]$.
The purities of the atomic and field states are then $\mathcal{P}_{atom}=2(p-1/2)^2+1/2$ and $\mathcal{P}_{field}=(1-\lambda)/(1+\lambda)$, respectively.
We take $p$ and $\lambda$ as independent control parameters of purities of the atom and the field.
The purity of the atom is 1 when $p=1$ while it shows the minimum value 1/2 when $p=1/2$,
and the purity of the field can be characterized by $0\leq\lambda\leq1$.
The initial states evolve through a JC interaction, $H_{JC} = g(\pjct{e}{g} a + \pjct{g}{e} a^\dagger)$,
where $g$ is the coupling strength and
$a$ ($a^\dagger$) is the annihilation (creation) operator of the field mode.
After the interaction with time $t$, we make a local projection on the field mode into a subspace spanned by $\ket{n}$ and $\ket{n+1}$, then the total state becomes
\begin{align}
&p\left(
\begin{array}{cccc}
P_{n-1} S_{n-1}^2 & 0 & 0 & 0 \\
0 & P_n C_n^2 & iP_n C_n S_n&0 \\
0 & -iP_n C_n S_n & P_n S_n^2 & 0 \\
0 & 0 & 0 & P_{n+1}C_{n+1}^2
\end{array}
\right) \nonumber \\
+&q\left(
\begin{array}{cccc}
P_n C_{n-1}^2 & 0 & 0 & 0 \\
0 & P_{n+1} S_{n+1}^2 & -iP_{n+1} C_n S_n&0 \\
0 & iP_{n+1} C_n S_n & P_{n+1} C_n^2 & 0 \\
0 & 0 & 0 & P_{n+2}S_{n+1}^2
\end{array}
\right),
\label{eq:BK}
\end{align}
where $C_n=\textnormal{cos}\left(gt\sqrt{n+1}\right)$, $S_n=\textnormal{sin}\left(gt\sqrt{n+1}\right)$, $P_n=(1-\lambda)\lambda^n$, and $q=1-p$.
We use negativity of partial transpose (NPT) as an entanglement measure \cite{NPT1,NPT2,NPT3}
  defined by $-2\textnormal{min}(0,\epsilon)$, where 
$\epsilon$ is the minimum eigenvalue of the partial transpose of the density matrix with respect to one of its parties.
Since the projection is a local operation, observing nonzero NPT of the projected density matrix indicates that the original total state has 
entanglement.

A nonzero value of NPT for  Eq.~(\ref{eq:BK}) with any value of $n$ is evidence of the atom-field entanglement.
Values of the NPT for several choices of $n$  with normalizations of  Eq.~(\ref{eq:BK})
   are presented in Fig.~\ref{fig:BK}.
If the atom was initially in a pure excited state ($p=1$), entanglement of the projected density matrix with $n=0$ shows entanglement except when $gt=n\pi/2$.
As shown in Fig.~\ref{fig:BK},
the cases with the other projections ($n=10$ and 100) of the density matrix lead to the conclusion that the entanglement always exists for $t>0$. This is in agreement with the result of Ref. \cite{BosePRL2001}.

 Figure~\ref{fig:BK} shows that the NPT tends to disappear as the purity of the atom decreases.
In  Fig.~\ref{fig:BK}, we also observe similar behaviors when the initial atom is in a maximally mixed
state $(\pjct{g}{g}+\pjct{e}{e})/2$ and the 
temperature of the field increases from 0 to infinity.
It seems that a certain degree of purity is required to generate entanglement. 
However, 
the method used here is merely to find a sufficient condition of the presence of entanglement, and
we cannot confirm that there is no entanglement in the original state when the NPT is zero for the projected density matrix \cite{BosePRL2001}.

\begin{figure}[t]
\includegraphics[width=245px]{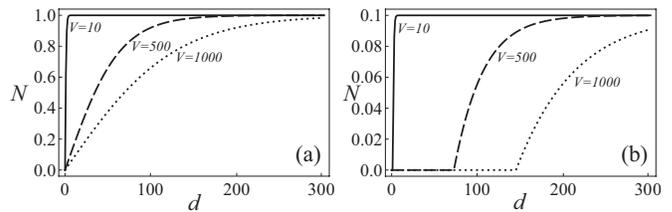}
\caption{NPT of the projected density matrix for Eq.~(\ref{eq:rho_gamma})  against displacement $d$ of the displaced thermal field. We choose amplitude $\gamma=2$ for the projection basis $\ket{\pm}_\gamma$ with several values of
$V$, and the normalized purity $r$ is (a)~1 and (b)~0.1. Unless the microscopic superposition is totally mixed as $r=0$, entanglement
is observed for $d\gg\sqrt{V}$ for any values of $V$. 
}
\label{fig:SeedThermal}
\end{figure}

We now consider an entanglement generation scheme between two harmonic oscillators
via a cross Kerr nonlinear interaction. This type of
interaction has been explored to study entanglement involving thermal states~\cite{JeongPRL2006,MauroPRA2006,JeongPRL2009,MK2010}.
It is analogous to the scheme in the previous example: the atomic state
is replaced by a superposition of the vacuum and the single photon, and the JC 
interaction is replaced by a cross-Kerr nonlinear interaction.
A limitation of the cross-Kerr interaction using fibers in implementing certain quantum gates 
\cite{Munro2005}
has been pointed out \cite{Shapiro2006,Shapiro2007,G2010}. However, 
we are here interested in possibility of entanglement generation using possible physical interactions in any systems. It 
was pointed out that a nonzero
conditional phase shift with a high fidelity is possible in a cross-Kerr interaction between pulses with unequal group velocities  \cite{He}.
Hosseini {\it et al.} demonstrated the feasibility of achieving a large cross-Kerr interaction 
at the single photon level based on a memory-based approach \cite{HoPreprint}.

The cross-Kerr interaction  between 
modes $a$ and $b$ is described by an interaction Hamiltonian $H_{\textnormal{Kerr}}=\chi a^\dagger a b^\dagger b$ where $\chi$ is the nonlinear interaction strength.
One mode is prepared as a superposition of the vacuum and the single photon state, generally in a mixture
\begin{equation}
\label{eq:Seed}
\rho_{\rm m}=\frac{1}{2}(|0\rangle\langle0|+r|0\rangle\langle1|+r|1\rangle\langle0|+|1\rangle\langle1|),
\end{equation}
where the purity, Tr[$\rho_{\textnormal{m}}^2$] = $(1+r^2)/2$, 
 is characterized by a real value $r$.
The other mode is prepared as a displaced thermal state, $\rho^{th}=\int d^2\alpha P^{th}_\alpha(V,d) |\alpha\rangle\langle\alpha|$, where $P^{th}_{\alpha}(V,d)=2\left[\pi(V-1)\right]^{-1}$exp$[-\frac{2|\alpha-d|^2}{V-1}]$ and $\ket{\alpha}$ is a coherent 
state of amplitude $\alpha$.
The variance $V$ 
is related to the average photon number $\bar{n}$ as $V=2(\bar{n}-d^2)+1$, and 
$d$ corresponds to the displacement of the state from the origin in phase space.
Note that the purity of $\rho^{th}$ is independent of $d$ since the displacement is a unitary operation.
After a cross-Kerr interaction  with an interaction time $t=\pi/\lambda$, the total state becomes
\begin{align}
\label{eq:SeedThermalInteraction}
\frac{1}{2}\big\{|0\rangle\langle0|\otimes\rho^{th}(V,d)+|1\rangle\langle1|\otimes\rho^{th}(V,-d)&\nonumber\\
+r|0\rangle\langle1|\otimes\sigma(V,d)+r|1\rangle\langle0|\otimes\sigma(V,-d)&\big\},
\end{align}
where $\sigma(V,d)=\int d^2\alpha P^{th}_\alpha |\alpha\rangle\langle\small{-}\alpha|$.
We make local projections on the total state into a subspace spanned by sets $\{|0\rangle,|1\rangle\}$ and $\{|+\rangle_\gamma,|-\rangle_\gamma\}$ 
where $|\pm\rangle_\gamma\equiv\mathcal{N}_\pm(|\gamma\rangle\pm|\small{-}\gamma\rangle)$ with $\mathcal{N}_\pm=1/\sqrt{2(1\pm 
\textnormal{exp}[-2\gamma^2])}$. 
The projected state $\rho_\gamma$ is then
\begin{align}
\label{eq:rho_gamma}
&\left(	
\begin{array}{cccc}
 \mathcal{N}^2_+C_\gamma & \mathcal{N}_+\mathcal{N}_{-}S_\gamma & r \mathcal{N}^2_+C_\gamma & -r\mathcal{N}_+\mathcal{N}_{-}S_\gamma\\ 
 \mathcal{N}_+\mathcal{N}_{-}S_\gamma & \mathcal{N}^2_-C_\gamma & r\mathcal{N}_+\mathcal{N}_{-}S_\gamma & -r\mathcal{N}^2_-C_\gamma \\ 
  r\mathcal{N}^2_+C_\gamma  & r\mathcal{N}_+\mathcal{N}_{-}S_\gamma & \mathcal{N}^2_+C_\gamma & -\mathcal{N}_+\mathcal{N}_{-}S_\gamma\\  
 -r\mathcal{N}_+\mathcal{N}_{-}S_\gamma & -r\mathcal{N}^2_-C_\gamma & -\mathcal{N}_+\mathcal{N}_{-}S_\gamma & \mathcal{N}^2_-C_\gamma
\end{array}
\right) \nonumber \\
&+R_\gamma\left(
\begin{array}{cccc}
 \mathcal{N}^2_+& \mathcal{N}_+\mathcal{N}_{-} & r \mathcal{N}^2_+ & -r\mathcal{N}_+\mathcal{N}_{-} \\ 
  \mathcal{N}_+\mathcal{N}_{-} & -\mathcal{N}^2_- & -r\mathcal{N}_+\mathcal{N}_{-} & r\mathcal{N}^2_-\\ 
  r\mathcal{N}^2_+ & -r\mathcal{N}_+\mathcal{N}_{-} & \mathcal{N}^2_+ & \mathcal{N}_+\mathcal{N}_{-}\\  
 -r\mathcal{N}_+\mathcal{N}_{-} & r\mathcal{N}^2_- & \mathcal{N}_+\mathcal{N}_{-} & -\mathcal{N}^2_-
\end{array}
\right),
\end{align}
where $C_\gamma =\frac{4}{V+1} \textnormal{exp}[-\frac{2}{V+1}(\gamma^2+d^2)]\textnormal{cosh}[4\gamma d/(V+1)]$, $S_\gamma =\frac{4}{V+1} 
\textnormal{exp}[-\frac{2}{V+1}(\gamma^2+d^2)]\textnormal{sinh}[4\gamma d/(V+1)]$,
and $R_\gamma = \frac{4}{V+1} \textnormal{exp}[-\frac{2}{V+1}(V\gamma^2+d^2)]$.
We assumed that $\gamma$ and $d$ are real without loss of generality.
Notice that $\textnormal{Tr}[\rho_\gamma(r)]\neq1$ because the local projection is not an unitary operation.
The absolute scale of the NPT is not important for the purpose of our study, because any nonzero value is meaningful enough.

\begin{figure}[t]
\includegraphics[width=245px]{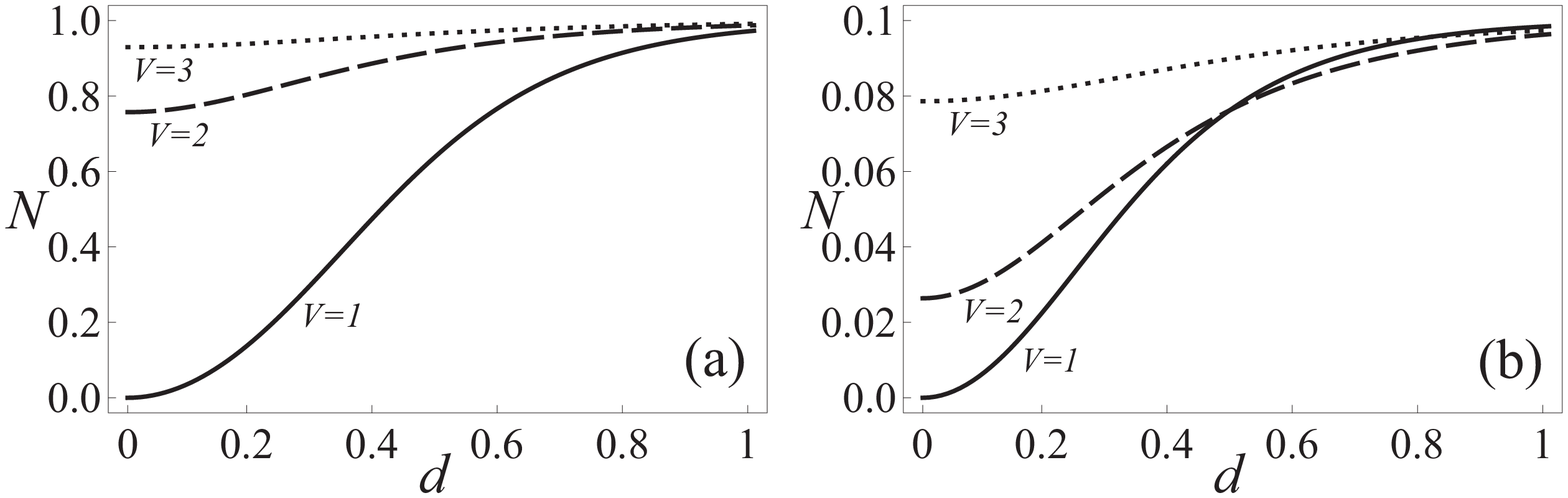}
\caption{NPT of the locally projected state of $\rho^{BS(+)}$ with (a)~$r=1$ and (b)~$r=0.1$. We used $\gamma=2$ for the projection basis. 
Entanglement is always observed for large values of $V\gg1$ even when $d=0$
unless $r=0$.}
\label{fig:BS}
\end{figure}

We numerically calculated the NPT of $\rho_\gamma (r)/\textnormal{Tr}[\rho_\gamma (r)]$ for some values of $V$, $d$, and $r$, and present the results in Fig.~\ref{fig:SeedThermal}.
When $d=0$, the NPT is  zero regardless of the value of $V$.
In this case, the original state is separable because $\sigma(V,0)^{T}=\sigma(V,0)$ 
in Eq.~(\ref{eq:SeedThermalInteraction}): the density matrix is then invariant under a partial transpose on one side.
However, we can observe entanglement for any values of $V$ with $d\gg\sqrt{V}$ as far as $r$ is nonzero in Eq.~(\ref{eq:Seed}).

As the next step, we consider two slightly different entanglement
generation schemes where conditioning measurements are used in addition to unitary interactions \cite{JeongPRL2006}.
After the Kerr interaction between a microscopic state in Eq.~(\ref{eq:Seed})  and a thermal state, we measure out the microscopic part on the basis 
$(|0\rangle \pm |1\rangle )/\sqrt{2}$ \cite{JeongPRL2006}.
The resultant state of the remaining mode is highly non-classical exhibiting singular behaviors on its Wigner function \cite{JeongPRL2006}.
After we transmit this state through a 50:50 beam splitter,
the state becomes
\begin{equation}
\begin{aligned}
&\rho^{BS(\pm)}=N\int d^2\alpha P^{th}_\alpha(V,d)
\big\{
|\delta\rangle\langle\delta| \otimes |\small{-}\delta\rangle\langle\small-\delta| + |\small{-}\delta\rangle\langle\small{-}\delta|  \\
&  \otimes |\delta\rangle\langle\delta|
\pm r |\delta\rangle\langle\small{-}\delta| \otimes |\small{-}\delta\rangle\langle\delta|
\pm r |\small{-}\delta\rangle\langle\delta| \otimes |\delta\rangle\langle\small{-}\delta| 
\big\},
\end{aligned}
\end{equation}
where $\delta=\alpha/\sqrt{2}$ and $N=(2\pm2r~\textnormal{exp}[-2d^2/V]/V)^{-1}$.
The NPT of the locally projected density matrix by sets $\{|+\rangle_\gamma,|-\rangle_\gamma\}$ for both parties is plotted in Fig.~\ref{fig:BS}.
One can observe that entanglement between thermal states is generated even when they are highly mixed ($V\gg1$) 
unless the initial microscopic state was totally mixed with $r=0$.
Interestingly, a high mixture ($V\gg1$) enables one to observe  entanglement even with $d=0$ and a small value of $r$.

\begin{figure}[t]
\includegraphics[width=245px]{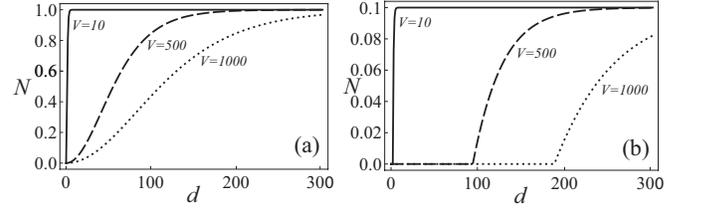}
\caption{NPT of the locally projected state of $\rho^{TT(+)}$ versus the displacement $d$ with (a)~$r=1$ and (b)~$r=0.1$. The amplitude of the projection basis is $\gamma=2$ as in Figs.~2 and 3. Each curve corresponds to $V=10 ($solid$), 500 ($dashed$),$ and $1000($dotted). Nonzero values of $d$ much larger than $\sqrt V$ are required to observe entanglement compared to 
$\rho^{BS(+)}$ cases. }
\label{fig:SeedTwoThermal}
\end{figure}

The other entanglement generation scheme between two macroscopic thermal states is as follows.
The microscopic state successively interacts with two thermal states via the cross-Kerr interactions, and it is measured out along the basis
$(|0\rangle \pm |1\rangle )/\sqrt{2}$. The resultant state is
\begin{align}
&\rho^{TT(\pm)}_{} =\nonumber\\
&N \big\{ \rho^{th}_a(V,d) \otimes \rho^{th}_b(V,d) + \rho^{th}_a(V,\small{-}d) \otimes \rho^{th}_b(V,\small{-}d)& \nonumber \\
&\pm r \sigma_a(V,d) \otimes \sigma_b(V,d)  \pm r \sigma_a(V,\small{-}d) \otimes \sigma_b(V,-d) \big\},
\end{align}
We also project this state into a subspace spanned by sets \{$|+\rangle_\gamma,|-\rangle_\gamma$\} for each mode, then calculate the NPT which is presented in Fig.~\ref{fig:SeedTwoThermal}.
Again, an initial microscopic state with any value of $r>0$ is useful to generate entanglement between thermal states regardless of $V$. Here, however, a condition of $d \gg \sqrt{V}$ is required to clearly observe entanglement.

\begin{figure}[t]
\includegraphics[width=180px]{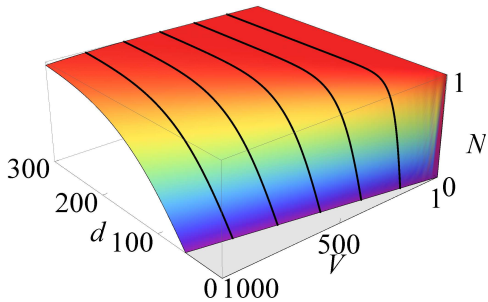}
\caption{(Color online) The NPT of $\rho^\psi_\gamma$ with $\gamma=2$. In this case $V$ is the only parameter that adjusts the purities of both parties simultaneously. We can 
always obtain the entanglement even in $V\rightarrow\infty$ 
limit with sufficient displacement $d\gg \sqrt V$. This indicates that nonzero purity of 
the system is sufficient to generate entanglement.}
\label{fig:TwoThermalKerr}
\end{figure}

We now come to a natural question: Can two highly mixed states, when their purities become arbitrarily small,
still be entangled through a direct unitary interaction?
Here we simply prepare two identical displaced thermal states, $\rho^{th}(V,d)$, in each mode then act cross-Kerr interaction directly between those states.
In order to calculate the effect of the interaction,
we first point out that two-mode states where each mode has a definite photon number parity are eigenstates of the cross-Kerr interaction for time $t=\pi/\chi$ so that $U=\textnormal{exp}[i\pi 
a^\dagger a b^\dagger b]$.
It works as a controlled-phase gate on the $\ket{\pm}_\gamma $ basis 
where $\ket{\pm}_\gamma$ is defined in the above  Eq.~(\ref{eq:rho_gamma})  \cite{CoherentCluster}:
$U\ket{+}_\gamma\ket{\pm}_\gamma =\ket{+}_\gamma\ket{\pm}_\gamma$
and
$U\ket{-}_\gamma\ket{\pm}_\gamma = \pm\ket{-}_\gamma\ket{\pm}_\gamma$.
Based on this, we calculate the evolution of the two thermal sates under $U$ as
\begin{equation}
U |\alpha\rangle|\beta\rangle
=\frac{1}{2}\big{(}|\alpha\rangle|\beta\rangle+|\alpha\rangle|\small{-}\beta\rangle+|\small{-}\alpha\rangle|\beta\rangle\small{-}|\small{-}\alpha
\rangle|\small{-}\beta\rangle\big{)}\equiv\ket{\psi}
\end{equation}
and
$\rho^\psi = \int d^2\alpha d^2\beta P^{th}_\alpha(V,d) P^{th}_\beta(V,d) |\psi\rangle\langle\psi|$.
The locally projected density matrix with basis $\{|+\rangle_\gamma,|-\rangle_\gamma\}$ is

\begin{align}
\rho^\psi_\gamma=\left(	
\begin{array}{cccc}
 X_\gamma^2 & X_\gamma S_\gamma & S_\gamma X_\gamma & -S_\gamma^2\\
 X_\gamma S_\gamma & X_\gamma Y_\gamma & S_\gamma^2 & -S_\gamma Y_\gamma\\
 S_\gamma X_\gamma & S_\gamma^2 & Y_\gamma X_\gamma & -Y_\gamma S_\gamma \\
 -S_\gamma^2 & -S_\gamma Y_\gamma & -Y_\gamma S_\gamma & Y_\gamma^2
\end{array}
\right)
\end{align}
where $X_\gamma=C_\gamma+R_\gamma$ and $Y_\gamma=C_\gamma-R_\gamma$ in terms of $C_\gamma$, $S_\gamma$, and $R_\gamma$ defined in  Eq.~(\ref{eq:rho_gamma}).
We plot the results in  Fig.~\ref{fig:TwoThermalKerr}, which shows that entanglement is generated even in the high temperature
 limit as long as $d\gg\sqrt V$.
We observe that two thermal states at arbitrarily high temperatures are entangled through a direct unitary interaction as long as the condition, $d\gg\sqrt{V}$, is satisfied.

In summary, we have studied several examples to explore
entanglement generation involving highly mixed physical systems.
Our results reveal some interesting facts concerning the
generation of entanglement involving such highly mixed systems.
The purity of initial states is not necessarily a prerequisite for
entanglement generation. It rather depends on the model
of the interaction between the initial states.
In particular, we have shown that entanglement between thermal states can be generated 
via a direct unitary interaction even when both states have purities arbitrarily close to zero.


This work was supported by the NRF grant funded by
the Korea government (MEST) (No. 3348-20100018)
and the World Class University program.
M.K. acknowledges support  from  the Global Ph.D. Fellowship of the NRF of Korea.
M.S.K. acknowledges the support by the
NPRP 4-554-1-084 from Qatar National Research Fund. M.S.K. is also grateful for the hospitality provided by the Seoul National University.

\end{document}